\documentclass[11pt,prd,aps,tightenlines,notitlepage,superscriptaddress,nofootinbib,preprintnumbers,letterpaper]{revtex4} 
\pdfoutput=1

\usepackage{amsmath, latexsym, amssymb, graphicx, ifpdf, slashed, color, hyperref, url, cancel, bbm, hyperref}
\usepackage{comment}
\usepackage[T1]{fontenc}
\usepackage[dvipsnames]{xcolor}
\hypersetup{colorlinks, citecolor=nicegreen, linkcolor=nicered, urlcolor=niceblue}
\definecolor{nicered}{rgb}{.7,.1,.1}
\definecolor{nicegreen}{rgb}{.2,.7,.1}
\definecolor{niceblue}{rgb}{0.1,0.2,0.6}
\definecolor{darkblue}{rgb}{0,0,.5}
\usepackage{orcidlink}
\usepackage{notes2bib}

\begin{document}

\title{
Entropy Dilution Mechanism and Dark Matter Isocurvature}

\author{Miha Nemev\v{s}ek \orcidlink{0000-0003-1110-342X} }
\email{miha.nemevsek@ijs.si}
\affiliation{Faculty of Mathematics and Physics, University of Ljubljana, 
Jadranska 19, 1000 Ljubljana, Slovenia}
\affiliation{Jo\v{z}ef Stefan Institute, Jamova 39, 1000 Ljubljana, Slovenia}

\author{Yue Zhang \orcidlink{0000-0002-1984-7450} }
\email{yzhang@physics.carleton.ca}
\affiliation{Department of Physics, Carleton University, Ottawa, ON K1S 5B6, Canada}


\begin{abstract}
  We explore dark matter isocurvature perturbations in cosmological scenarios with entropy 
  dilution resulting from a late decaying massive state that temporarily dominates the universe 
  before big-bang nucleosynthesis.
  The dilution mechanism features a non-standard scaling window of the radiation energy 
  density, which allows for a new way to derive the entropy dilution factor.
  On super-horizon scales, the same scaling window also governs the evolution of 
  isocurvature, and we present an analytic approach to solving the perturbation equations.
  We characterize various regimes where dark matter isocurvature survives and
  is subject to constraints from the cosmic microwave background observations.
  For cosmologically viable scenarios, we find that dark matter is dominantly produced 
  from its dilutor decay, and the isocurvature is suppressed by the entropy dilution factor
  to the $4/3$ power.
  Our work demonstrates the complementarity between isocurvature and collisionless damping 
  of dark matter as powerful probes of the entropy dilution mechanism.
\end{abstract}

\maketitle


\section{Introduction}

A compelling theory of dark matter must satisfy two primary criteria. 
It should offer a calculable production mechanism in the early universe that 
explains the origin of the observed dark matter relic abundance.
The mechanism should be controlled by a set of parameters tied to some 
experimental observables that would make the theory testable.

The best example of this is the Weakly interacting massive particle (WIMP), 
which undergoes thermal freeze out and recognizes that the electroweak scale is  
special in the sense that it produces the correct dark matter relic. 
It is the most explored dark matter candidate, with numerous limits derived from 
(in)direct detection and collider experiments~\cite{Cirelli:2024ssz}. 
The relic density explanation fails if the WIMP mass moves below the GeV scale, 
known as the Lee-Weinberg bound~\cite{Lee:1977ua}. 
Essentially, a smaller mass reduces the annihilation cross section and lifts the 
Boltzmann suppression in the dark matter number density, leading to overproduction.
This interplay has motivated a recent quest for light dark forces between dark matter 
and Standard Model (SM) particles, where the dark matter mass can be in the sub-GeV 
range with new direct detection prospects~\cite{Essig:2022dfa}.
The above overproduction argument stops working for sufficiently light dark matter 
that freezes out while still being ultra-relativistic. 
In this regime, reducing the dark matter mass results in a lower relic density. 
In fact, the SM neutrinos with the sum of their masses of around 11~eV are consistent 
with the relic density requirement. 
Nevertheless, such a possibility is ruled out due to other observations, including the
large scale structure of the universe~\cite{Hu:1997mj, Lesgourgues:2006nd, Wong:2011ip} 
(hot dark matter constraint) and the KATRIN experiment~\cite{KATRIN:2024cdt} 
(absolute neutrino mass constraint).

The leading solution to avoid the problem of hot dark matter is late entropy injection 
from the decay of a non-relativistic particle species whose energy density temporarily 
dominates the universe before the decay~\cite{Scherrer:1984fd}. 
Here ``late'' states that the decay must occur after the relativistic freeze out of 
dark matter. 
On the other hand, in order to fulfill the constraints on energy injection and the 
equation of state of the universe for successful big-bang nucleosynthesis~\cite{Kawasaki:2004qu, 
An:2023buh}, the decay needs to take place at temperatures above MeV.
Due to the entropy injection, it takes more time for the universe to cool down to 2.7 K today, 
thereby providing extra cooling time for the dark matter particles to pass the structure 
formation constraint. 
We will denote dark matter particles as $X$, and use $Y$ to denote the decaying particle 
that plays the role of entropy dilution (the dilutor). 
The KATRIN constraint can be simply avoided if $X$ is a different particle from the 
SM neutrinos.
A number of model building works have been done to embed the entropy dilution mechanism 
in various well-motivated beyond-SM theories~\cite{Moroi:1999zb, Asaka:2006ek, Bezrukov:2009th, 
Acharya:2009zt, Nemevsek:2012cd, Zhang:2015era, Patwardhan:2015kga, Berlin:2016vnh, Soni:2017nlm, 
Evans:2019jcs, Dror:2020jzy, Asadi:2021bxp, Nemevsek:2022anh, Nemevsek:2023yjl, Bleau:2023fsj, Guo:2024mar}.

The cooling of dark matter via the entropy dilution mechanism also helps to alleviate 
the dark matter overproduction problem. 
Indeed, consider a generic warm dark matter candidate whose phase space distribution is 
thermal and characterized by its own temperature $T_X$. 
Its number density is proportional to $T_X^3$. 
To comprise 100\% of dark matter, $T_X$ is related to dark matter mass by
\begin{equation}
  T_X \simeq 0.09 T_\gamma \left( \frac{6.5\,\rm keV}{m_X} \right)^{1/3} \, ,
\end{equation}
where $T_\gamma$ is the cosmic microwave background (CMB) photon temperature, and 6.5 keV 
is the lower bound on warm dark matter set by the latest small-scale structure 
observations~\cite{DES:2020fxi}.
The smallness of $T_X/T_\gamma$ is made possible if the dilutor $Y$ dominantly decays 
into SM particles.
The dilution factor $\mathcal{S}$ is defined as the entropy ratio in radiation after
and before $Y$ domination and is often estimated using the sudden decay approximation
in the literature.
In section~\ref{sec:dilutionfactor}, we present a new way to derive $\mathcal{S}$ by
defining a non-standard scaling window of the radiation energy density.

The dark matter dilution mechanism makes novel predictions that can be tested (mainly) 
with cosmological data.\footnote{In case the dilutor decays within a macroscopic distance 
comparable to beam dump experiments, such as SHIP~\cite{Alekhin:2015byh} or other LLP 
searches~\cite{Alimena:2019zri}, one may be able to test the microscopic origin of cosmic
dilution in the laboratory.}
If the outcome of dilution is warm dark matter, the dark matter power spectrum features
a collisionless damping on sub-Mpc length scales, which is the target of small scale
structure observations, including the Lyman-$\alpha$ forest, strong gravitational lensing, 
and Milky Way dwarf galaxies~\cite{Enzi:2020ieg, Nadler:2021dft}.
Furthermore, the significance of the dilutor to dark matter partial decay ($Y\to X$) has 
been investigated in~\cite{Nemevsek:2022anh}.
This decay channel produces very energetic dark matter particles whose free-streaming effect 
can affect large-scale structure observations of the matter power spectrum. 
An upper bound on the branching ratio, ${\rm Br}_{Y\to X} \lesssim 1 \%$, has been found 
using the Sloan Digital Sky Survey (SDSS) data~\cite{Reid:2009xm} by assuming equal and
thermal initial abundances of $X$ and $Y$.
Detailed implications of this result have been explored in the context of the left-right 
symmetric model, which also addresses the origin of neutrino mass~\cite{Nemevsek:2023yjl}.

In this work, we go beyond the assumption made previously by allowing for different initial 
abundances of $X$ and $Y$ and discuss how the structure formation constraint changes with 
this new moving part.
More importantly, we study another potential signature of the dilution mechanism, which lies 
in dark matter isocurvature. 
In section~\ref{sec:isocurvature}, we follow a gauge invariant description of the density 
perturbations (usually denoted as $\zeta$~\cite{Bardeen:1983qw, Wands:2000dp}) 
and allow them to differ on super-horizon scales for $X$, $Y$, and the SM radiation.
This is possible if each sector inherits quantum fluctuations from their own scalar field 
during inflation~\cite{Mukhanov:1981xt, Linde:1984ti, Sasaki:1986hm, Mukhanov:1988jd}.~\footnote{
If $X$ were once in thermal equilibrium with the SM plasma, their super-horizon perturbations 
would be set equal. 
But the perturbations in the dilutor $Y$, if not thermalized, can be very different.
For other works on dark matter production from the decay of curvaton-like particles,
see~\cite{Gerlach:2025uxo, Sainio:2009vg}.}
Similar assumptions were made for the curvaton scenario as the origin of cosmological 
perturbations~\cite{Lyth:2001nq, Moroi:2001ct}. 
However, the entropy dilution scenario considered here differs from the curvaton in 
several aspects.
First of all, our dilutor $Y$ need not be the curvaton or even a scalar particle, and 
second, we insist that $Y$ must temporarily dominate the universe for significant entropy 
injection to occur ($\mathcal{S} \gg 1$).
In our scenarios, the dark matter relic density always has two components: a population of $X$ 
that already existed before $Y$ domination, and the secondary $X$ particles from $Y$ decay. 
Finally, we allow the initial perturbations from different sectors to be comparable in magnitude, 
and the stringent isocurvature constraints are set on the entropy dilution parameters.

We develop a semi-analytic approach to solve the evolution of various $\zeta$s throughout 
the course of $Y$ decay. 
In particular, we show that the same entropy dilution factor $\mathcal{S}$, relevant for 
the dark matter relic density, is also the key parameter governing how efficiently different 
perturbations evolve into one another.
We carefully evolve the dark matter perturbations from early times when $X$ particles are 
produced relativistically to late times when they all become cold, and the horizon crossing
of perturbation modes is encoded in CMB observations.
We project constraints on dark matter isocurvature set by the Planck space 
observatory~\cite{Planck:2018jri} on the general dark matter dilution parameter space. 
In section~\ref{sec:other}, we showcase the complementarity between isocurvature and other 
cosmological probes.
Our main results are summarized in Fig.~\ref{fig:moneyplot}.

\section{Overview of the Entropy Dilution Mechanism} \label{sec:dilutionfactor}

In this section, we work with the unperturbed Friedmann–Lema\^{i}tre–Robertson–Walker (FLRW) 
universe and discuss aspects of the entropy dilution mechanism for producing the observed 
dark matter relic density.
This includes the period of temporary matter domination, a non-standard scaling window 
for the energy density of the thermal plasma, and a new approach to calculating the 
dilution factor.

\begin{figure}[h]
  \begin{center}
    \includegraphics[width=0.5\textwidth]{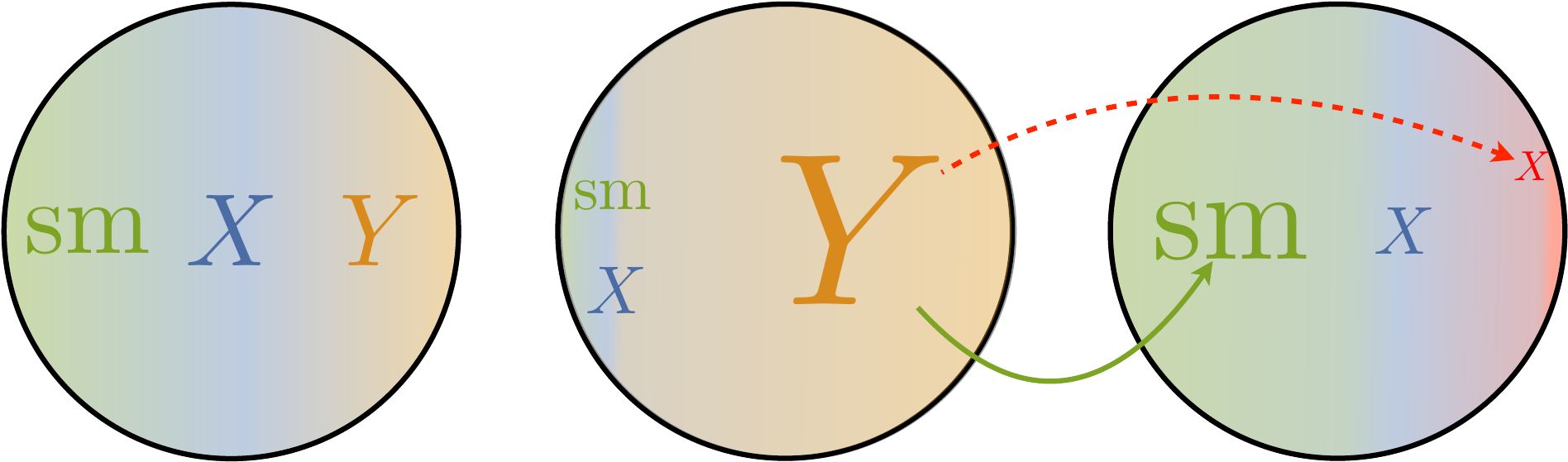}
  \end{center}
  \caption{Schematic illustration of the DM dilution mechanism: relativistic 
  freeze-out, dilutor matter domination, entropy production (solid green) and 
  DM re-population (dashed red).}
\label{fig:Illust}
\end{figure}

We start by setting up the problem on very general grounds. 
A schematic cartoon of dilution is shown in Fig.~\ref{fig:Illust}. 
The early universe is comprised of three species that have decoupled from each other: the 
SM thermal plasma, dark matter $X$, which we assume is still made of relativistic particles, 
and the diluting particles $Y$, which have already become non-relativistic. 
The abundances of $X$ and $Y$ are given by their yields
\begin{align}
  Y_X &= \frac{n_X}{s} \, , & 
  Y_Y &= \frac{n_Y}{s} \, , &
  s   &= \frac{2\pi^2}{45} g_*(T) T^3 \, ,
\end{align}
defined as the ratio of their initial number densities to the entropy density of the 
thermal plasma at times that are sufficiently early, such that the decay of $Y$ has 
not started yet.
Had this universe evolved without $Y$, the relic abundance $X$ today would be given by
\begin{equation}\label{eq:OmegaXwithoutY}
  \Omega_X^0 = \frac{m_X Y_X s_0}{\rho_c} \, ,
\end{equation}
where the entropy density today is $s_0 = 2891.2\,{\rm cm}^{-3}$ and the critical energy 
density is $\rho_c = 1.054\times10^{-5}h^2\,{\rm GeV cm^{-3}}$. 
We work in the parameter space where the above $\Omega_X > 0.12 \, h^{-2}$ and $X$ would 
overclose the universe. 
This easily occurs if dark matter $X$ is a thermal relic that froze out 
relativistically from the thermal plasma. 
With a number density similar to that of the SM neutrinos, $X$ would over-close the 
universe if its mass were above $\sim 11$ eV, which is too hot to produce the 
observed structures in the universe~\cite{White:1983fcs}.
This gives strong motivation to consider a setup with a dilutor $Y$, whose decay 
supplies a late time entropy injection and reduces the DM yield $Y_X$.

The dilutor $Y$ is in the form of matter, and if it is sufficiently long-lived, it could 
temporarily dominate the total energy density of the universe, as illustrated in Fig.~\ref{fig:Illust}.
We assume this occurs throughout this work so that there is a significant amount 
of entropy release.
To fulfill the goal of entropy dilution, the main decay channel of $Y$ must be into 
the SM particles $Y \to \text{SM}$.
We also consider the possibility of
\begin{equation}\label{eq:DecayYtoX}
  Y \to n \, X + m \, {\rm SM} \, , 
\end{equation}
as a subdominant decay channel, with a branching ratio ${\rm Br}_{Y\to X} \ll 1$,
where $n$ and $m$ are the multiplicities of $X$ and SM particles per $Y$ decay, respectively.
The secondary $X$ produced in this way also contributes to the final dark matter relic density.

\subsection{Sudden Decay Approximation}

The sudden decay approximation has been widely used in various dark matter models for relic 
density calculations, where the entropy dilution factor $\cal S$ is estimated as follows.
We assume that the universe was matter ($Y$) dominated before $t_*$, when all the $Y$ 
particles suddenly decay
\begin{equation} \label{eq:SDA}
  \frac{2}{3\tau_Y} = H(t_*) \simeq \sqrt{\frac{8\pi \bar \rho_Y(t_*)}{3 M_{\rm pl}^2}} \, .
\end{equation}
The energy density of $Y$ shortly before $t_*$ is 
\begin{equation}
  \bar \rho_Y(t_*) = m_Y Y_Y s_< \, .
\end{equation}
Immediately after the decay of $Y$, the universe becomes radiation dominated, with the 
Hubble parameter evolving continuously.
If we allow for a small fraction of the dilutor $Y$ to decay into $X$, the radiation energy 
density has to be reduced appropriately
\begin{equation}
  \bar\rho_R(t_*) = \bar\rho_Y(t_*) \left( 1 - y \, {\rm Br}_{Y \to X} \right) \, .
\end{equation}
Here, $y$ is the energy fraction carried by $X$ particle(s) in the second $Y$ decay channel
in Eq.~\eqref{eq:DecayYtoX}.
The corresponding entropy density of radiation immediately after $t_*$ can be determined
because the SM plasma is back to thermal equilibrium with a temperature $T_*$
\begin{align}\label{eq:ThermalRelations}
  \bar \rho_R(t_*) &= \frac{\pi^2}{30} g_*(T_*) T_*^4 \, , 
  & 
  s_> &= s_R(t_*) = \frac{2\pi^2}{45} g_*(T_*)T_*^3 \simeq 
  1.01 \, g_{*}(T_*)^{\frac{1}{4}}\bar\rho_R(t_*)^{\frac{3}{4}} \, ,
\end{align}
and we can solve for $T_*$ in terms of $\tau_Y$ using Eq.~\eqref{eq:SDA}.
The equations above allow us to derive the entropy densities before and after the decay 
of $Y$ as
\begin{align}
  s_< &= \frac{M_{\rm pl}^2}{6\pi m_Y Y_Y \tau_Y^2} \, ,
  &
  s_> &= \frac{2 \left( 1 - y \, {\rm Br}_{Y \to X} \right)^{\frac{3}{4}} 
  g_*(T_*)^{\frac{1}{4}} M_{\rm pl}^{\frac{3}{2}}}{9 (5\pi)^{\frac{1}{4}} \tau_Y^{\frac{3}{2}}} \, .
\end{align}
The resulting entropy dilution factor is then given by
\begin{equation}\label{eq:DilutionFactorSDA}
  \mathcal{S}_{\text{SDA}} = \frac{s_>}{s_<} \simeq 2.10 \, 
  \left( 1 - y \, {\rm Br}_{Y\to X} \right)^{\frac{3}{4}} g_*(T_*)^{\frac{1}{4}} 
  m_Y Y_Y \sqrt\frac{\tau_Y}{M_{\rm pl}} \, .
\end{equation}

The final relic density of dark matter $X$ today can be rescaled from 
Eq.~\eqref{eq:OmegaXwithoutY} by dividing by the dilution factor
\begin{equation}\label{eq:OmegaXwithY0}
\begin{split}
  \Omega_X h^2 = \frac{\Omega_X^0 h^2}{\mathcal{S}} 
  \left( 1 + \frac{Y_Y}{Y_X} n \, {\rm Br}_{Y\to X}\right) \, .
\end{split}
\end{equation}
Throughout this work, we assume that all the $X$ particles in the universe remain collisionless
after their primordial abundance (and perturbations) are fixed initially, and their number
density redshifts as $a^{-3}$ with the expansion of the universe.
The bracket takes into account the secondary $X$ particles from $Y$ decay as part 
of the final relic density.

\subsection{Beyond Sudden Decay}\label{sec:beyondSDA}

\begin{figure}[t]
  \begin{center}
    \includegraphics[width=0.618\textwidth]{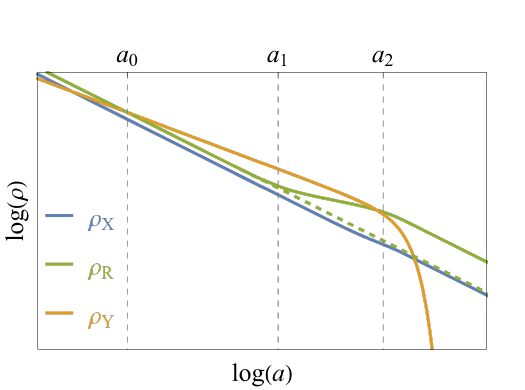}
  \end{center}
  \caption{
  Schematic plot of the energy density evolution in entropy dilution via the 
  out-of-equilibrium decay of a ``long-lived'' particle $Y$. 
  The blue, green and orange curves stand for the energy density of dark matter 
  $X$, the thermal plasma $R$, and dilutor $Y$, respectively.
  The three special scale factors $a_{0, 1, 2}$ are shown by the vertical gray 
  dashed lines.
  The temporary matter domination of $Y$ commences at a scale factor $a = a_0$.
  The non-standard scaling window discussed in sections~\ref{sec:beyondSDA} 
  and~\ref{sec:scaling} corresponds to $a$ between $a_1$ and $a_2$, when significant 
  entropy production takes place and $a_2$ corresponds to the time when $Y$ decays.
  }
\label{fig:rho}
\end{figure}

In this work, we go beyond the simple sudden decay approximation by taking into account the 
period of time it takes for $Y$ to first dominate the energy density and decay.
The evolution of energy densities is governed by the set of equations 
\begin{align} \label{eq:RhoYEvol}
  \bar\rho_Y' &= -3 \bar\rho_Y - \frac{1}{H\tau_Y} \bar\rho_Y \, , 
  \\ \label{eq:RhoREvol}
  \bar\rho_R' &= -\left(4 + \frac{g_*'}{3g_*} \right) \bar\rho_R + 
  \frac{1 - y {\rm Br}_{Y\to X}}{H\tau_Y} \bar\rho_Y \, , 
  \\ \label{eq:RhoXEvol}
  \bar\rho_X' &= - 4 \bar\rho_X + \frac{y {\rm Br}_{Y\to X}}{H\tau_Y} \bar\rho_Y \, ,
\end{align}
where the ${\,}^\prime$ stands for $\text{d}/\text{d} \log a$, $a$ is the scale 
factor of the universe, $\Gamma_Y = 1/\tau_Y$, and the $y \in [0,1]$ parameter 
denotes the fraction of energy carried by the $X$ species per $Y$ decay through
Eq.~\eqref{eq:DecayYtoX}~\cite{Nemevsek:2022anh, Nemevsek:2023yjl}.
The Hubble parameter is given by the Friedmann equation
\begin{equation}
  H^2 = \frac{8\pi}{3 M_{\rm pl}^2} \left( \bar\rho_Y + \bar\rho_R + \bar\rho_X \right) \, .
\end{equation}
A typical numerical solution to the above equations is illustrated in Fig.~\ref{fig:rho}.
The energy density of the thermal plasma $\bar\rho_R$ evolves along the solid green curve.
If $Y$ did not exist, $\bar \rho_R$ would simply scale as $a^{-4}$ and would follow 
the dashed green line.
The entropy dilution factor $\cal S$ can be inferred by comparing the values of $\bar\rho_R$ 
on the solid and dashed curves at a time sufficiently long after $Y$ decay, using the 
thermodynamic relations in Eq.~\eqref{eq:ThermalRelations}.

\subsection{Non-standard Scaling Window of $\bar\rho_R$}\label{sec:scaling}

A notable feature of the $\bar \rho_R$ evolution in Fig.~\ref{fig:rho} is the presence 
of an intermediate window~\cite{Scherrer:1984fd, Chung:1998rq, Giudice:2000ex}, where 
$\bar \rho_R$ redshifts with a power law that is slower than the usual $a^{-4}$. 
The window begins at scale factor $a_1$ and ends at a scale factor $a_2$, as shown by 
the vertical dashed lines in Fig.~\ref{fig:rho}. 
The physical origin of this scaling is due to $Y$ matter dominating the energy 
budget and then enforcing its scaling onto $\bar \rho_R$ by being its dominant
source. 
This is analogous to any externally driven process, such as an oscillator, which 
inherits the time dependence of the dominating external source.
We can understand it analytically by examining Eq.~\eqref{eq:RhoYEvol} 
and~\eqref{eq:RhoREvol}, which also gives us a new way of deriving the entropy dilution 
factor $\cal S$.

At early times, well before the lifetime of $Y$ that coincides with $a_2$, the $\bar\rho_Y$
in~\eqref{eq:RhoYEvol} scales as
\begin{equation}\label{eq:RhoYSolution}
  \bar \rho_Y (a) 
  \propto a^{-3} \, .
\end{equation}
Throughout the entire $Y$ matter dominated era $a \in [a_0, a_2]$, the Hubble parameter 
scales as
\begin{equation}\label{eq:Hscaling}
  H \propto a^{-\frac{3}{2}} \, .
\end{equation}
As a result, the last term of Eq.~\eqref{eq:RhoREvol}, which is proportional to 
$\bar \rho_Y/H$, scales as $a^{-3/2}$.
The onset of the new scaling window occurs at the scale factor $a_1$ when this 
term dominates over the $-4 \bar \rho_R$ term on the right-hand side. 
Neglecting the $g_*'$, Eq.~\eqref{eq:RhoREvol} admits a simple power scaling 
solution for radiation energy density for $a \in [a_1, a_2]$,
\begin{equation}\label{eq:ScalingWindow0}
  \bar\rho_R \propto a^{-\frac{3}{2}} \, ,
\end{equation}
and more precisely
\begin{equation}\label{eq:ScalingWindow}
  \bar\rho_R = \frac{2 \left( 1 - y {\rm Br}_{Y\to X} \right) \bar \rho_Y}{5 H \tau_Y} \, .
\end{equation}
This scaling window is in sharp contrast with the regular evolution of $\bar\rho_R$ 
due to the expansion of the universe with $\bar\rho_R \propto a^{-4}$, which occurs 
in the absence of sources, regardless of whether we are in a radiation- or 
matter-dominated universe.
The new scaling window ends at $a_2$ once $Y$ decays
\begin{equation} \label{eq:defa2}
  t_2 = \frac{2}{3H (a_2)} = \tau_Y \, ,
\end{equation}
where the factor of $2/3$ is needed for the time-Hubble relation during the matter 
($Y$) dominated universe, similar to the one in Eq.~\eqref{eq:SDA}.

\subsection{A New Derivation of Dilution Factor}

To derive the entropy dilution factor in the above picture, we need to determine 
the scale factor $a_1$ and $a_2$ in terms of the model parameters.
It is useful to introduce a scale factor $a_1 - \varepsilon$ corresponding to a 
time slightly before the above scaling window, when the universe has already entered 
the $Y$ dominated era.
The initial energy densities for $Y$ and radiation at this time are 
$\bar\rho_Y(a_1 - \varepsilon)$ and $\bar\rho_R(a_1 - \varepsilon)$, respectively.
We will take the $\varepsilon \to 0$ limit in the end.

During $a_1 - \varepsilon < a < a_2$, the Hubble parameter is given by
\begin{equation}
  H^2 \simeq \frac{8\pi \bar\rho_Y}{3 M_{\rm pl}^2} \, ,
\end{equation}
with $\bar\rho_Y \propto a^{-3}$, and the energy density of radiation $\bar\rho_R \propto a^{-4}$.
Extrapolating $\bar \rho_R$, $\bar \rho_Y$, and $H$ from their values at $a_1 - \varepsilon$ to $a_1$,
where they begin to fulfill the scaling condition Eq.~\eqref{eq:ScalingWindow}, we obtain
\begin{equation}\label{eq:a1}
  \frac{a_1-\varepsilon}{a_1} = \left[ \frac{2(1-y {\rm Br}_{Y\to X})}{
  5 H(a_1-\varepsilon)\tau_Y} \frac{\bar \rho_Y(a_1-\varepsilon)}{\bar\rho_R(a_1-\varepsilon)}
  \right]^{\frac{2}{5}} \, .
\end{equation}
On the other hand, using Eq.~\eqref{eq:defa2}, we get
\begin{equation}\label{eq:a2}
  \frac{a_1-\varepsilon}{a_2} = \left[ \frac{2}{3H(a_1-\varepsilon)\tau_Y} \right]^{\frac{2}{3}} \, .
\end{equation}

Taking the ratio leads to
\begin{equation}\label{eq:a1a2Ratio}
  \frac{a_2}{a_1} = 1.21 (1-y {\rm Br}_{Y\to X})^{\frac{2}{5}} g_*(a_1)^{\frac{2}{15}} 
  \left(\frac{m_Y^2 Y_Y^2 \tau_Y}{M_{\rm pl}}\right)^{\frac{4}{15}} \, ,
\end{equation}
where we used $\bar\rho_Y = m_Y Y_Y s$, applied the thermal relation 
$\bar\rho_R \simeq 0.99 s^{4/3}g_*^{-1/3}$, and took the $\varepsilon \to 0$.

For the scaling window to exist, we need $a_2 > a_1$, which requires $Y$ to be sufficiently 
abundant to begin with
\begin{equation}
  Y_Y \gtrsim \sqrt{\frac{M_{\rm pl}}{m_Y^2 \tau_Y}} \, ,
\end{equation}
in the small ${\rm Br}_{Y\to X}$ limit.
Compared to the dilution factor given in Eq.~\eqref{eq:DilutionFactorSDA}, the above lower 
bound on $Y_Y$ corresponds to requiring $\mathcal{S}>1$.
Thus, whenever significant entropy dilution is to take place, the $\bar\rho_R$ scaling 
window must exist in the first place.
As we will see below, identifying the relevant scaling $[a_1, a_2]$ window not only yields an 
elegant derivation of the entropy dilution factor but also allows for analytically solving the 
large-scale isocurvature perturbations.

Let us derive the entropy dilution factor by evolving the thermal plasma from $a_1$ to 
$a_2$, where $\bar\rho_R$ satisfies the scaling solution, Eq.~\eqref{eq:ScalingWindow0}.
This allows us to write the entropy density at $a_2$ in terms of the radiation energy 
density at $a_1$
\begin{equation}
  s(a_2) \simeq 1.01 \, g_*(a_2)^{\frac{1}{4}} \bar\rho_R(a_2)^{\frac{3}{4}} 
  \simeq 1.01 \, g_*(a_2)^{\frac{1}{4}} \bar\rho_R(a_1)^{\frac{3}{4}} 
  \left( \frac{a_1}{a_2} \right)^{\frac{9}{8}} \, .
\end{equation}
In contrast, if the dilution effect from $Y$ did not exist, the entropy density would 
simply redshift as $a^{-3}$ between $a_1$ and $a_2$, leading to
\begin{equation}
  \tilde s(a_2) \simeq 1.01 \, g_*(a_1)^{\frac{1}{4}} \bar\rho_R(a_1)^{\frac{3}{4}} 
  \left( \frac{a_1}{a_2} \right)^{3} \ll s(a_2) \, .
\end{equation}
The entropy dilution factor is 
\begin{equation}\label{eq:EntropyDilutionFactor}
  \mathcal{S} \equiv \frac{s(a_2)}{\tilde s(a_2)} = \left( \frac{g_*(a_2)}{g_*(a_1)} 
  \right)^{\frac{1}{4}} \left( \frac{a_2}{a_1} \right)^{\frac{15}{8}} \, .
\end{equation}
Plugging in the ratio found in Eq.~\eqref{eq:a1a2Ratio}, we get
\begin{equation}
  \mathcal{S} \simeq 1.43 \, (1-y {\rm Br}_{Y\to X})^{\frac{3}{4}} g_*(a_2)^{\frac{1}{4}} 
  m_Y Y_Y\sqrt{\frac{\tau_Y}{M_{\rm pl}}} \simeq \frac{2}{3} \mathcal{S}_{\text{SDA}} \, .
\end{equation}
This result has the same parametric dependence as the dilution factor derived in the 
sudden decay approximation, Eq.~\eqref{eq:DilutionFactorSDA}, but is smaller by a factor 
of $\sim 1.5$.

In the small ${\rm Br}_{Y\to X}$ limit, we have
\begin{equation}\label{eq:Smoreaccurate}
  \mathcal{S} \simeq 1.43 \, g_*(a_2)^{\frac{1}{4}} m_Y Y_Y \sqrt{\frac{\tau_Y}{M_{\rm pl}}} \, .
\end{equation}
In the numerical analysis of~\cite{Scherrer:1984fd}, a slightly higher prefactor of 1.8 was 
obtained, which takes into account the decay of the remnant $Y$ population at times $t>\tau_Y$.

As mentioned earlier, we assume that all the $X$ particles are collisionless after production.
Using Eqs.~\eqref{eq:OmegaXwithoutY}, \eqref{eq:OmegaXwithY0}, and the dilution factor derived 
in Eq.~\eqref{eq:Smoreaccurate}, the final dark matter relic density is given by
\begin{equation}\label{eq:OmegaXwithY}
\begin{split}
    \Omega_X h^2 = 0.12\ g_*(T_2)^{-\frac{1}{4}} \left(\frac{Y_X}{Y_Y}+ n {\rm Br}_{Y\to X}\right) 
    \left( \frac{4.5\times10^6 m_X}{m_Y} \right) \sqrt{\frac{1\,\rm sec}{\tau_Y}} \ ,
\end{split}
\end{equation}
where $T_2$ is the photon temperature at the scale factor $a_2$. 
It is equal to $T_*$ used in the sudden decay approximation.

The final dark matter relic is comprised of two parts: the initial population of $X$ 
that exists before $Y$ decay and the secondary ones that are produced from $Y$ decay.
Their fractions are controlled by the ratio of $Y_X/Y_Y$ versus $n {\rm Br}_{Y\to X}$.
Eq.~\eqref{eq:OmegaXwithY} implies that the entropy dilution effect is more efficient with 
heavier and longer-lived $Y$.
Because $\tau_Y$ is constrained by BBN to be shorter than $\sim 1$ sec, the dilutor $Y$ 
must be at least a million times heavier than dark matter $X$ if they have similar 
initial abundances ($Y_X \simeq Y_Y$), say from thermal relativistic freeze-out.

\section{Dilution with Isocurvature Initial Perturbations}\label{sec:isocurvature}

In this section, we explore density perturbations in the FRW universe. 
We work in the conformal Newtonian gauge, where the perturbed FRW metric tensor takes the 
usual form 
\begin{equation}
  \text{d} s^2 = -\left(1 + 2 \psi \right) 
  \text{d}t^2 + a(t)^2 \left(1 - 2 \phi \right) \text{d} \vec{x}\cdot \text{d} \vec{x} \, .
\end{equation}
The Bardeen potentials $\psi(\vec{x}, t), \phi(\vec{x}, t)$ are space-time-dependent 
perturbations of the metric that satisfy the perturbed Einstein equations
\begin{equation}\label{eq:Einstein}
  3 H \dot\phi + 3 H^2 \psi + \frac{k^2 \phi}{a^2} = - 4\pi G \sum_\alpha \delta \rho_\alpha  \, .
\end{equation}
For the rest of our discussion, we neglect small anisotropic stress perturbations in the 
energy-momentum tensor and set $\psi = \phi$.

The energy density perturbations $\delta \rho_\alpha(\vec{x}, t)$ evolve on top of the leading-order 
average energy densities $\bar\rho_\alpha(t)$, which are defined in the homogeneous and isotropic limit.
The species index $\alpha$ goes through dark matter $X$, dilutor $Y$, and the SM thermal plasma $R$.
Our discussion will first cover the stage where $Y$ has already become non-relativistic, but the $X$ 
particles are still relativistic.
Thus, the equations of state $w_\alpha = P_\alpha/\rho_\alpha$ for each of the components in the plasma
are $w_X = w_R = 1/3$ and $w_Y = 0$.
In a later part of this section, we will discuss the transition of $w_X$ from $1/3$ to $0$, 
which is assumed to take place at much later times.

\subsection{Super-Horizon Modes}

We are most interested in isocurvature perturbations relevant for observations of wavelengths 
that are much larger than the size of the horizon during $Y$ decay. 
The average energy densities and perturbations evolve according to the continuity 
equations~\cite{Weinberg:2004kr}
\begin{equation}\label{eq:RhoDeltaRhoEvolution}
\begin{split}
  \dot{\bar\rho}_\alpha + 3 H (\bar P_\alpha + \bar\rho_\alpha) &= Q_\alpha \, , 
  \\
  \delta\dot{\rho}_\alpha + 3 H(\delta \rho_\alpha + \delta P_\alpha) - 3 \dot\phi (\bar P_\alpha 
  + \bar\rho_\alpha) &= Q_\alpha \psi + \delta Q_\alpha \, ,
\end{split}
\end{equation}
where the over dot stands for $\text{d}/\text{d}t$ and $H = \dot a/a$. 
The second equation is valid for super-horizon modes with the fluid velocity and gradient 
terms neglected.
The source terms on the right-hand side are nonzero due to the decay of $Y$ into the SM 
particles and $X$. 
They take the following form
\begin{equation}\label{eq:QandDeltaQ}
\begin{array}{ll}
  Q_Y = - \bar\rho_Y/\tau_Y \ , &\hspace{2cm} \delta Q_Y = - \delta\rho_Y/\tau_Y \, , 
  \\
  Q_R = (1 - y {\rm Br}_{Y\to X}) \bar\rho_Y/\tau_Y \ , &\hspace{2cm} 
  \delta Q_R = (1 - y {\rm Br}_{Y\to X}) \delta\rho_Y/\tau_Y \, , 
  \\
  Q_X = y {\rm Br}_{Y\to X} \bar\rho_Y/\tau_Y \ , &\hspace{2cm} 
  \delta Q_X = y {\rm Br}_{Y\to X} \delta\rho_Y/\tau_Y \, . 
\end{array}
\end{equation}
The first equation of \eqref{eq:RhoDeltaRhoEvolution} is consistent with 
Eqs.~\eqref{eq:RhoYEvol}--\eqref{eq:RhoXEvol}.

It is useful to define the gauge invariant quantity for each species
\begin{equation}
  \zeta_\alpha = -\psi - H \frac{\delta \rho_\alpha}{\dot{\bar\rho}_\alpha} \, .
\end{equation}
Using Eqs.~\eqref{eq:Einstein} and \eqref{eq:RhoDeltaRhoEvolution}, it follows that the 
evolution of each $\zeta_\alpha$ is governed by~\cite{Malik:2004tf, Malik:2002jb}
\begin{equation}\label{eq:ZetaAlpha}
  \dot \zeta_\alpha = \frac{Q_\alpha \dot H}{\dot{\bar\rho}_\alpha} \left( \frac{\delta \rho}{\dot{\bar\rho}} - \frac{\delta \rho_\alpha}{\dot{\bar\rho}_\alpha} \right) - \frac{H}{\dot{\bar\rho}_\alpha} \left( \delta Q_\alpha - \frac{\dot Q_\alpha}{\dot{\bar\rho}_\alpha} \delta \rho_\alpha \right) + \frac{3H^2}{\dot{\bar\rho}_\alpha} \left( \delta P_\alpha - \frac{\dot{\bar P}_\alpha}{\dot{\bar\rho}_\alpha} \delta \rho_\alpha \right) \ ,
\end{equation}
where $\bar\rho = \sum_\alpha \bar\rho_\alpha$ and $\delta\rho = \sum_\alpha \delta \rho_\alpha$. 
As already mentioned, we focus on super-horizon perturbation modes with wavelengths much larger 
than the horizon when the dilution mechanism takes place and neglect terms suppressed by $k^2/H^2$.
The same equation holds for all super-horizon modes.
The last term on the right-hand side corresponds to the intrinsic non-adiabatic pressure and also 
vanishes by neglecting the gradient of $\delta P_\alpha$ and $\delta \rho_\alpha$.

In the absence of source terms $Q_\alpha, \delta Q_\alpha$, all terms on the right-hand side vanish.
In this case, each $\zeta_\alpha$ is conserved when the perturbation mode is outside the horizon.
The initial conditions for $\zeta_\alpha$ likely originate from quantum fluctuations that occur 
during inflation. 
In general, it is plausible to imagine additional scalar field(s) besides the inflaton, and their 
quantum fluctuations are produced independently.
After the end of inflation, if they decay into $X$, $Y$, and the SM sector separately, the 
resulting $\zeta_X$, $\zeta_Y$, and $\zeta_R$ generically differ from each other. 
The differences serve as the isocurvature initial conditions for Eq.~\eqref{eq:ZetaAlpha}.

In some earlier works~\cite{Bezrukov:2009th, Nemevsek:2012cd, Dror:2020jzy, Nemevsek:2022anh, 
Nemevsek:2023yjl}, the entropy dilution mechanism was embedded in gauge extensions of the SM.
$X$ and $Y$ are identified as right-handed neutrinos that are in thermal equilibrium with the 
Standard Model particles at high temperatures through right-handed current interactions.
The thermal contact strongly suppresses the differences among $\zeta_\alpha$, even if they were 
set differently immediately after inflation, as discussed on general grounds in~\cite{Weinberg:2004kf}.
In this case, the initial condition for Eq.~\eqref{eq:ZetaAlpha} is adiabatic. 
In this work, we will not limit ourselves to such a simplified scenario, but explore the 
implications of isocurvature perturbations and observational constraints for the entropy 
dilution mechanism.

To calculate the evolution of $\zeta_\alpha$ in the dilution mechanism, we plug 
Eq.~\eqref{eq:QandDeltaQ} into the right-hand side of \eqref{eq:ZetaAlpha} to get
\begin{align}
  \zeta_Y' &= - \frac{H' \bar \rho_Y}{\tau_Y H^2 \bar\rho_Y'} (\zeta_Y - \zeta) \, , \label{eq:zetaY} 
  \\
  \zeta_R' &= \frac{(1-y{\rm Br}) {\bar\rho}_Y'}{\tau_Y H{\bar\rho}_R'} \left[ \frac{H' \bar \rho_Y}{H {\bar\rho}_Y'} 
  \left(\zeta_R - \zeta \right) - \left(\zeta_R - \zeta_Y \right) \right] \, , \label{eq:zetaR} 
  \\
  \zeta_X' &= \frac{y{\rm Br} {\bar\rho}_Y'}{\tau_Y H{\bar\rho}_X'} \left[ \frac{H' \bar \rho_Y}{H {\bar\rho}_Y'} 
  \left(\zeta_X - \zeta\right) - \left(\zeta_X - \zeta_Y\right) \right] \, , \label{eq:zetaX}
\end{align}
where the $^\prime$ stands for $d/d\log a = (1/H)d/dt$, and
\begin{equation}\label{eq:ZetaTotal}
  \zeta \equiv \frac{1}{\dot{\bar\rho}}\sum_\alpha \dot{\bar\rho}_\alpha \zeta_\alpha = 
  -\psi -H \frac{\delta\rho}{\dot{\bar\rho}} \, .
\end{equation}

Eqs.~\eqref{eq:zetaY}--\eqref{eq:zetaX} are linear equations for $\zeta_\alpha$ and can be 
readily solved with the knowledge of the leading order quantities $\bar\rho_\alpha$ and $H$, 
which is the task of the previous section.
It is obvious that with an adiabatic initial condition where all $\zeta_\alpha$ are equal, 
all the terms on the right-hand side vanish. 
In this case, $\zeta_X \simeq \zeta_Y \simeq \zeta_R = \zeta$ is a conserved quantity outside 
the horizon, and $\zeta$ is equal to the curvature perturbation $\mathcal{R}$~\cite{Weinberg:2003sw}.

On the other hand, when $\zeta_X$, $\zeta_Y$, and $\zeta_R$ have different initial values, the 
super-horizon modes evolve with the expansion of the universe.
For $\zeta_R$ and $\zeta_X$, because the right-hand sides of their equations are proportional 
to $\bar\rho_Y'$, their evolutions are tied to the decay of the dilutor $Y$.
After $Y$ decays away and vanishes from existence in the universe, $\zeta_R$ and $\zeta_X$ will 
freeze-in and remain conserved afterwards.
It is also worth noting that in the presence of isocurvature, $\zeta$ is no longer a conserved 
quantity.
For example, during the temporary $Y$ dominated era, Eq.~\eqref{eq:ZetaTotal} implies 
$\zeta \simeq \zeta_Y$, whereas early on, if the universe was radiation dominated, 
$\zeta \simeq \zeta_R$.
A linear combination of Eqs.~\eqref{eq:zetaY}--\eqref{eq:zetaX} gives (see appendix~\ref{app:zetatot} 
for the derivation)
\begin{equation}\label{eq:ZetaTotalEvolve}
  \zeta' = - \frac{\bar\rho_Y'}{\bar\rho'} (\zeta - \zeta_Y) \ .
\end{equation}
Because the coefficient of the bracket $\bar\rho_Y'/\bar\rho'$ is positive definite, $\zeta$ 
is attracted towards $\zeta_Y$ if they have different values to start with.
This attractor effect is most important during the $Y$ dominated era.

\subsection{Approximate Analytic Solutions}

\begin{figure}[t]
  \begin{center}
    \includegraphics[width=1\textwidth]{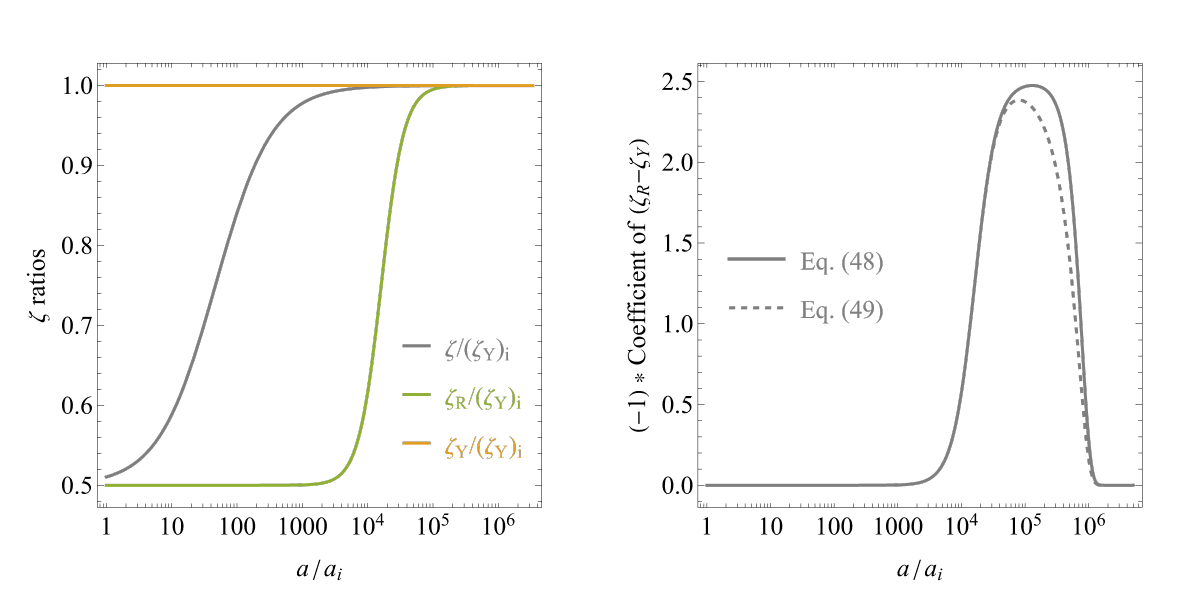}
  \end{center}
  \caption{
  {\it Left panel:} evolution of $\zeta, \zeta_R, \zeta_Y$ in a dark matter dilution scenario, 
  for a super-horizon mode. 
  We set parameters $m_Y = 100\,$GeV, $\tau_Y = 0.1\,$sec and work in the ${\rm Br}_{Y\to X}\ll1$ 
  limit. 
  For the initial conditions, $a_i$ corresponds to a temperature of the radiation plasma $T_i=m_Y$. 
  The $Y$ particles are already non-relativistic with initial abundance $Y_Y=0.0235$.
  The initial perturbations are $\zeta_{Y,i}/\zeta_{R,i}=2$. 
  We solve the $\zeta$ equations~\eqref{eq:zetaY}, \eqref{eq:zetaR} and \eqref{eq:ZetaTotalEvolve} 
  from $a_i$ to late times after $Y$ has decayed away.
  {\it Right panel:} Coefficients of $(\zeta_R-\zeta_Y)$ on the right-hand side of the $\zeta_R$ 
  equations, \eqref{eq:zetafluidsApprox} and \eqref{eq:zetafluidsApproxFurther}. Within the $[a_1,a_2]$ 
  scaling window, the coefficient is approximately equal to $-5/2$. 
  See Eq.~\eqref{eq:5/2}.}\label{fig:zetaR}
\end{figure}

In this subsection, we show that, under a few good approximations, the solutions to the 
$\zeta_\alpha$ evolution equations can be derived analytically.
The first observation we make by comparing Eqs.~\eqref{eq:zetaY}--\eqref{eq:zetaX} 
and~\eqref{eq:ZetaTotalEvolve} is that the right-hand side of the $\zeta'$ equation is not 
proportional to $\Gamma_Y/H$, which is a small factor before $Y$ decay begins.
This implies that $\zeta$ will start to evolve (toward $\zeta_Y$) as soon as $Y$ comes into 
domination and well before any other $\zeta_\alpha$ has time to change.
This behavior is clearly shown in the left panel of Fig.~\ref{fig:zetaR}.

During the $Y$ dominated era, $\bar\rho_Y'\simeq \bar\rho'$, and Eq.~\eqref{eq:ZetaTotalEvolve} 
is approximately
\begin{equation}
  \zeta' \simeq \zeta_Y - \zeta \, .
\end{equation}
This equation is to be contrasted with the $\zeta_Y$ evolution equation \eqref{eq:zetaY}, 
which takes the approximate form during $Y$ domination
\begin{equation}
  \zeta_Y' \simeq - \frac{1}{2 \tau_Y H} (\zeta_Y - \zeta) \, .
\end{equation}
At early times, before the lifetime of $Y$, we have $H^{-1} \ll \tau_Y$, and therefore approximately
\begin{equation}
  \frac{d(\zeta_Y-\zeta)}{d\log a} \simeq - (\zeta_Y - \zeta) \, .
\end{equation}
The difference between the total $\zeta$ and the dilutor $\zeta_Y$ evolves with the expansion 
of the universe as
\begin{equation}
  \left|\zeta - \zeta_Y\right| \sim \frac{1}{a} \, .
\end{equation}
This solution implies that $\zeta$ quickly approaches the dilutor's $\zeta_Y$ after a few 
decades of expansion, whereas the corresponding change in $\zeta_Y$ is highly suppressed. 

In the following discussion of $\zeta_{R, X}$ evolutions, we will make the approximation that 
\begin{equation}
  \zeta \simeq \zeta_Y \simeq \zeta_{Y, i} \, ,
\end{equation}
where $\zeta_{Y, i}$ denotes the initial perturbation of the dilutor $Y$.

Under this approximation, the $\zeta_R$ equation in \eqref{eq:zetaR} now takes the 
approximate form
\begin{equation}\label{eq:zetafluidsApprox}
\begin{split}
  \zeta_R' &\simeq \frac{(1-y{\rm Br}) {\bar\rho}_Y'}{\tau_Y H{\bar\rho}_R'} 
  \left( \frac{H' \bar \rho_Y}{H {\bar\rho}_Y'} - 1\right) \left(\zeta_R - \zeta_Y \right) \, .
\end{split}
\end{equation}
With $\zeta \simeq \zeta_Y$, we neglect the further evolution of $\zeta_Y$ and effectively treat 
it as a constant.
Through Eq.~\eqref{eq:zetafluidsApprox}, $\zeta_{R, X}$ are also attracted to $\zeta_Y$.
The right-hand side is most important at times smaller than $\tau_Y$ (see the scale factor 
$a_2$, defined in Eq.~\eqref{eq:a2}).
For $a \ll a_2$, we can approximate the $\bar\rho_Y$ solution in Eq.~\eqref{eq:RhoYSolution} as 
$\bar\rho_Y'\simeq -3 \bar\rho_Y$, and correspondingly, the Hubble parameter satisfies 
$H'\simeq -\frac{3}{2}H$.
With this, the big bracket factor in Eq.~\eqref{eq:zetafluidsApprox} is approximately $-1/2$.
The $\zeta_R$ equation further simplifies to
\begin{equation}\label{eq:zetafluidsApproxFurther}
\begin{split}
  \zeta_R' &\simeq \frac{3(1-y{\rm Br}) {\bar\rho}_Y}{2\tau_Y H{\bar\rho}_R'} 
  \left(\zeta_R - \zeta_Y \right) \, . 
\end{split}
\end{equation}
On the right-hand side, the coefficient outside the bracket is closely related to the scaling 
window discussed in section~\ref{sec:scaling}.
The scale factor $a_1$ defines the onset of the scaling window.
For $a\ll a_1$, $\bar\rho_R' \simeq -4 \bar\rho_R$, and ${\bar\rho}_R \gg (1-y{\rm Br}) 
{\bar\rho}_Y/(\tau_Y H)$. 
As a result, the coefficient of $(\zeta_R - \zeta_Y)$ is highly suppressed.
This suppression then gets lifted once the scaling window is reached, $a_1 < a < a_2$. 
Using Eq.~\eqref{eq:ScalingWindow}, we have 
${\bar\rho}_R'\simeq - \frac{3}{2} \bar\rho_R \simeq -\frac{3}{5}(1-y{\rm Br}) {\bar\rho}_Y/(\tau_Y H)$, 
and
\begin{align}\label{eq:5/2}
  \zeta_R' &\simeq -\frac{5}{2} \left(\zeta_R - \zeta_Y \right) \, ,
  & a_1 &< a <a_2 \, .
\end{align}
The $[a_1, a_2]$ window for $\zeta_R$ evolution and the coefficient $5/2$ are shown 
in the right panel of Fig.~\ref{fig:zetaR}.
Solving the above equation gives the final value of $|\zeta_R-\zeta_Y|$
\begin{equation}
  \zeta_{R,f} - \zeta_{Y,i} \simeq (\zeta_{R,i} - \zeta_{Y,i}) 
  \left( \frac{a_2}{a_1} \right)^{-\frac{5}{2}} \, .
\end{equation}
Hereafter, the lower indices $i, f$ denote the corresponding $\zeta$ values before and after 
the course of $Y$ decay.
In the above discussions, the differences among $\zeta_R, \zeta_Y, \zeta$ are power-law 
suppressed with the expansion of the universe. 
Interestingly, this differs significantly from the case of thermal freeze-out relics, where 
the corresponding $\zeta$ difference is exponentially suppressed~\cite{Weinberg:2004kf}.

It is useful to recall the entropy dilution factor $\mathcal{S}$ derived in 
Eq.~\eqref{eq:EntropyDilutionFactor}. 
Barring the change of $g_*$ before and after $Y$ decay (to the one-fourth power), we derive 
a connection between the evolution of $\zeta_R$ and the dilution factor
\begin{equation}\label{eq:zetaRY}
  \zeta_{R,f} - \zeta_{Y,i} \simeq \frac{\zeta_{R,i} - \zeta_{Y,i}}{\mathcal{S}^{\frac{4}{3}} } \, .
\end{equation}
This is one of our central results.
In the large $\mathcal{S}$ limit, the right-hand side is suppressed, rendering
$\zeta_{R,f} \simeq \zeta_{Y,f} \simeq \zeta_{Y,i}$.

While the model and cosmological parameters could vary, the evolutionary behavior discussed above
and shown in Fig.~\ref{fig:zetaR} remains robust as long as $Y$ domination takes place for a
sufficiently extended period.
The main point is that $\zeta_Y$ barely changes; the total $\zeta$ is attracted to $\zeta_Y$ 
long before $\zeta_R$ starts to evolve.
$\zeta_R$ mainly evolves during the scaling window defined in section~\ref{sec:scaling}, and 
finally, all the $\zeta$ evolutions shut off after $Y$ decays away.
Remarkably, even though the entropy dilution mechanism has a number of moving parts (model 
and cosmological parameters), the evolution of $\zeta_R$ is entirely governed by the dilution factor.

\subsection{Evolution of Dark Matter Isocurvature}

In the entropy dilution mechanism that we consider here, the final dark matter relic density is 
comprised of two components, including the initial population, which already exists before $Y$ decay 
(denoted by $X_1$) and the secondary population from $Y$ decay (denoted by $X_2$).
The total $\zeta_X$ is given by
\begin{equation}
  \zeta_X = \frac{\dot{\bar\rho}_{X_1} \zeta_{X_1} + \dot{\bar\rho}_{X_2} \zeta_{X_2}}{\dot{\bar\rho}_{X_1} 
  + \dot{\bar\rho}_{X_2}} \, .
\end{equation}
During the production of $X_2$ particles, their average energy density $\bar\rho_{X_2}$ evolves 
according to Eq.~\eqref{eq:RhoXEvol}.
The equation for $\bar\rho_{X_1}$ is similar but has no source term.
The evolution of $\zeta_{X_2}$ is given by Eq.~\eqref{eq:zetaX}, and the solution is 
$\zeta_{X_2} \simeq \zeta \simeq \zeta_Y \simeq \zeta_{Y,i}$.
This can be understood because all the $X_2$ particles are the decay products of $Y$.
On the other hand, the equation for $\zeta_{X_1}$ is sourceless; thus, $\zeta_{X_1}$ is conserved 
and equal to its initial value $\zeta_{X_1} = \zeta_{X, i}$.~\footnote{If the pre-existing 
$X$ particles are produced by freezing out relativistically from the thermal plasma, their 
initial perturbations $\zeta_{X,i}$ and $\zeta_{R,i}$ will be equal. 
This assumption could be generalized as well.}
This allows us to write
\begin{equation}\label{eq:zetaXtot}
  \zeta_X = \frac{\dot{\bar\rho}_{X_1} \zeta_{X,i} + \dot{\bar\rho}_{X_2} \zeta_{Y, i}}{\dot{\bar\rho}_{X_1} 
  + \dot{\bar\rho}_{X_2}} \, .
\end{equation}
The time evolution of $\zeta_X$ is thus entirely governed by the average energy densities.

\begin{figure}[t]
  \begin{center}
    \includegraphics[width=1\textwidth]{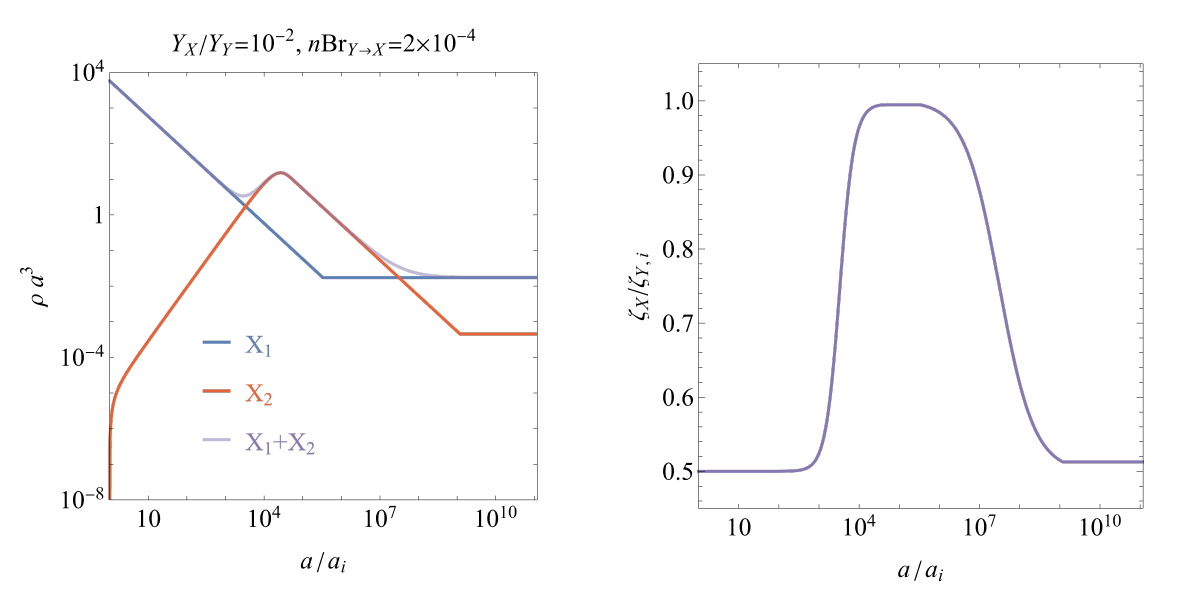}\\
    \includegraphics[width=1\textwidth]{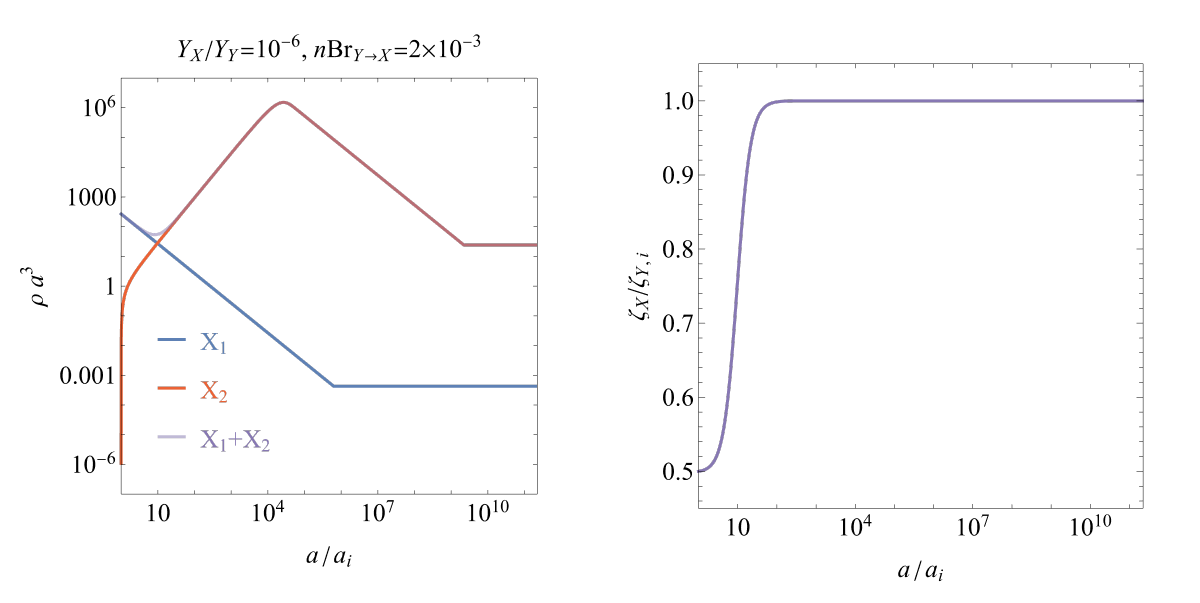}
  \end{center}
  \caption{
  Time evolution of dark matter energy density and isocurvature, for a super-horizon mode, and for
  two sets of parameters listed in Table~\ref{tab:param}.
  In both cases, the initial perturbations are set to satisfy $\zeta_{Y,i}/\zeta_{X,i}=2$.
  The initial scale factor $a_i$ corresponds to photon temperature $T_i=m_Y$. 
  In the {\it first row}, the energy density of secondary $X$ from $Y$ decay temporarily dominates 
  over the pre-existing ones ($X_1$). 
  This drives the total $\zeta_X$ to $\zeta_{Y,i}$ during the period of $\bar\rho_{X_2}>\bar\rho_{X_1}$, 
  as shown in the right-panel. 
  However, the secondary $X$ particles are more energetic and stay relativistic for a 
  longer period than the primary population.
  At late times, $\bar\rho_{X_2}$ becomes subdominant and $\zeta_X$ returns to $\zeta_{X,i}$. 
  In the {\it second row}, the secondary $X$ is more abundant and $\bar\rho_{X_2}$ always stays 
  higher than $\bar\rho_{X_1}$. 
  As a result, $\zeta_X$ is attracted to $\zeta_{Y,i}$ and remains conserved afterwards.}
  \label{fig:zetaX}
\end{figure}

\begin{table}[h]
  \centering 
  \begin{tabular}{c|c|c|c|c|c|c|c}
  \hline
  $m_X$ (keV) & $m_Y$ (GeV) & $\tau_Y$ (sec) & ${\rm Br}_{Y\to X}$ & $n$  & 
  $Y_Y$ & $Y_X/Y_Y$ & $\mathcal{S}$ 
  \\
  \hline
  $30$ & $10$ & $10^{-2}$ & $10^{-4}$ & 2 & 3.88 & $10^{-2}$ & 2782 
  \\
  $160$ & $100$ & $10^{-4}$ & $10^{-3}$ & 2 & 2.35 & $10^{-6}$ & 1767 
  \\
  \hline
  \end{tabular}
  \caption{Model and cosmological parameters in the first (second) row are used for 
  making the figures in the first (second) row of Fig.~\ref{fig:zetaX}.
  The parameters are chosen to produce the observed total dark matter relic density 
  via Eq.~\eqref{eq:OmegaXwithY}. We assume $Y\to XX$ as the second partial decay 
  channel in Eq.~\eqref{eq:DecayYtoX} so that $n=2$. 
  Note the $Y_Y$ used here are well above the thermal values, which is permitted if the 
  dilutor $Y$ has never equilibrated with the SM plasma.
  The dilution factor $\mathcal{S}$ is computed using Eq.~\eqref{eq:Smoreaccurate}.
  }
 \label{tab:param}
\end{table}

Throughout this work, we make the assumption that the $X$ particles are non-interacting 
after production.
They rely on the expansion of the universe to become non-relativistic and serve as dark matter 
in the universe today.
It is important to note that $\zeta_X$ can still evolve after $Y$ decays away, simply because the 
two dark matter components $X_1, X_2$ transition from radiation to matter ($w = 1/3 \to 0$) 
at different times.~\footnote{In Appendix~\ref{app:zetaconservation} we show that 
$\zeta_{X_1}, \zeta_{X_2}$ are conserved during the transition.}
Due to the large mass hierarchy between $m_X$ and $m_Y$, needed to accommodate the dark matter 
relic density (see Eq.~\eqref{eq:OmegaXwithY}), the $X_2$ particles are likely more energetic
and make the transition later than $X_1$.
This allows one to envision a non-trivial scenario where $\dot{\bar\rho}_{X_2}\gg\dot{\bar\rho}_{X_1}$
when they were both relativistic, but $\dot{\bar\rho}_{X_2}\ll\dot{\bar\rho}_{X_1}$ today, as 
shown in the upper left panel of Fig.~\ref{fig:zetaX}.
In this case, Eq.~\eqref{eq:zetaXtot} tells us that $\zeta_X$ is temporarily given by 
$\zeta_{Y_i}$ immediately after $Y$ decay, but later returns to $\zeta_{X_i}$, as shown by the
upper right panel of Fig.~\ref{fig:zetaX}.
On the other hand, if $\dot{\bar\rho}_{X_2}$ remains larger than $\dot{\bar\rho}_{X_1}$, 
the value of $\zeta_X$ will stay equal to $\zeta_{Y_i}$.
This is the story told in the second row of Fig.~\ref{fig:zetaX}.

The measurement of dark matter isocurvature~\cite{Planck:2018jri} takes place around the time 
when the CMB is formed.
Assuming all the $X_{1,2}$ particles have turned sufficiently non-relativistic by this 
time~\footnote{If this was not the case, there would be additional constraints. 
They will be discussed in the next section.}, 
their continuity equations read $\dot{\bar\rho}_{X_{1,2}} = -3 H {\bar\rho}_{X_{1,2}}$.
The late-time value of $\zeta_X$ can be written in terms of the fractions of dark matter relic 
density in Eq.~\eqref{eq:OmegaXwithY}
\begin{equation}\label{eq:zetaXY}
  \zeta_{X,f} = \frac{\bar\rho_{X_1} \zeta_{X_i} + \bar\rho_{X_2} \zeta_{Y,i}}{
  \bar\rho_{X_1} + \bar\rho_{X_2}} = f_1 \zeta_{X,i} + f_2 \zeta_{Y,i} \, ,
\end{equation}
where the fractions $f_{1,2}$ are defined as
\begin{align}\label{eq:f1f2}
  f_1 &\equiv \frac{Y_X/Y_Y}{Y_X/Y_Y+n{\rm Br}_{Y\to X}} \, , &
  f_2 &\equiv \frac{n{\rm Br}_{Y\to X}}{Y_X/Y_Y+n{\rm Br}_{Y\to X}} \, , &
  f_1 + f_2 &= 1 \, .
\end{align}

\subsection{CMB Constraint on Dark Matter Isocurvature}\label{sec:isocurvatureconstraint}

In momentum space, the dark matter isocurvature is given by
\begin{equation}
  S_{XR} = 3 \left[\zeta_{X,f} - \zeta_{R,f} \rule{0mm}{3.5mm}\right] 
  =-3H \left( \frac{\delta \rho_X}{\dot{\bar\rho}_X} - \frac{\delta \rho_R}{
  \dot{\bar\rho}_R} \right) \, .
\end{equation}
If all the $X$ particles have already turned non-relativistic, one can use the continuity 
equations to derive $S_{XR} = \delta \rho_X/\bar\rho_X - 3 \delta \rho_R/(4 \bar\rho_R)$.

We are interested in the large-scale perturbation modes that are relevant for the CMB 
observations made by Planck. 
These modes are well outside the horizon during $Y$ decay which must occur before big-bang 
nucleosynthesis, and it justifies the super-horizon analysis above.
After $Y$ has decayed away and before horizon entry, $S_{XR}$ remains a conserved quantity.
In the above discussions, we have suppressed the co-moving momentum $k$ dependence in 
$\zeta_\alpha$, $(\alpha = X, Y, R)$.
The initial condition for $\zeta_\alpha$ is of course ${\bf k}$ (or ${\bf r}$) dependent 
and randomly fluctuates in Hubble patches separated by distances $\gtrsim k^{-1}$, with the 
average $\langle \zeta_\alpha \rangle=0$.
In Eqs.~\eqref{eq:zetaRY} and \eqref{eq:zetaXY} which describes the evolution of super-horizon 
modes, the dilution factor, the yields, and branching ratio are all $k$ independent.
Therefore, the $k$ dependence of $S_{XR}$ is inherited directly from the initial perturbations 
$\zeta_{\alpha,i}({\bf k})$.

\begin{figure}[t]
  \begin{center}
    \includegraphics[width=0.618\textwidth]{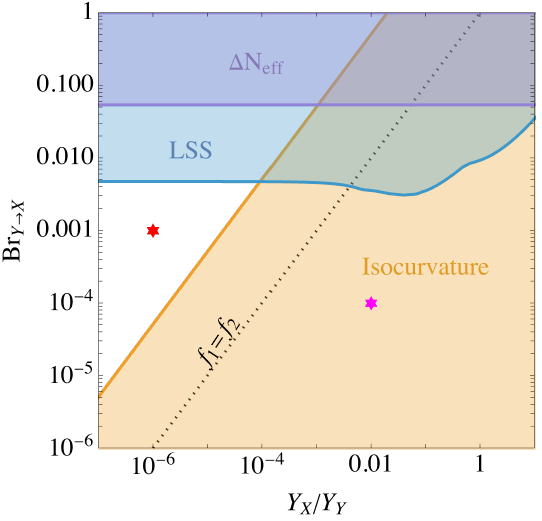}
  \end{center}
  \caption{
  Constraints on the key parameters of the dark matter dilution mechanism, dilutor to dark matter 
  decay branching ratio ${\rm Br}_{Y\to X}$ versus the ratio of initial abundances $Y_Y/Y_X$. 
  We consider $Y\to XX$ decay as the decay channel in Eq.~\eqref{eq:DecayYtoX}, so that $n=2$, 
  $m=0$ and $y=1$. 
  For the initial density perturbations, we assume $P_{YY}=4P_{XX}$ which is consistent with the 
  assumptions in Fig.~\ref{fig:zetaX} and $P_{XY}=0$.
  We work in the large dilution factor limit ($\mathcal{S}\gg1$).
  The orange shaded region is excluded by the Planck measurement of dark matter isocurvature, 
  as discussed in section~\ref{sec:isocurvatureconstraint}.
  The blue shaded region is excluded by the secondary dark matter free-streaming effect 
  discussed in section~\ref{sec:freestreaming}.
  The purple shaded region is excluded for giving too large $\Delta N_{\rm eff}$ during 
  the BBN epoch, as discussed in section~\ref{sec:neff}.
  The black dashed line corresponds to equal amount of initial and secondary $X$ populations 
  ($f_1=f_2$), where $f_{1,2}$ are related to ${\rm Br}_{Y\to X}$ and $Y_Y/Y_X$ via Eq.~\eqref{eq:f1f2}.
  The red and magenta stars correspond to the benchmark parameters listed in 
  Table~\ref{tab:param} and used for Fig.~\ref{fig:zetaX}. }
  \label{fig:moneyplot}
\end{figure}

The power spectra of the adiabatic and isocurvature perturbations are
\begin{equation}
\begin{split}
   \left\langle \zeta_{R,f}({\bf k}) \zeta_{R,f}({\bf k'})  \right\rangle &\equiv (2\pi)^3\delta_D({\bf k}-
   {\bf k'})\mathcal{P}_{\mathcal{R}\mathcal{R}}(k) \ , 
   \\ 
   \left\langle S_{XR}({\bf k}) S_{XR}({\bf k'}) \right\rangle &\equiv (2\pi)^3\delta_D({\bf k}-
   {\bf k'})\mathcal{P}_{\mathcal{I}\mathcal{I}}(k) \ , 
   \\
   \left\langle \zeta_{R,f}({\bf k}) S_{XR}({\bf k'}) \right\rangle &\equiv (2\pi)^3\delta_D({\bf k}-
   {\bf k'})\mathcal{P}_{\mathcal{R}\mathcal{I}}(k) \ ,
\end{split}
\end{equation}
where $\delta_D$ is the Dirac delta function, and the angle bracket means the 
ensemble average.
For the observed range of $k$, $P_{\mathcal{R}\mathcal{R}}$ must be nearly scale 
invariant with the measured tilt parameter $n_s \simeq 0.96$.
The Planck constraint on isocurvature is set on two parameters~\cite{Planck:2018jri}
\begin{align}
  \beta_{\rm iso} &= \frac{\mathcal{P}_{\mathcal{II}}}{\mathcal{P}_{\mathcal{RR}} + 
  \mathcal{P}_{\mathcal{II}}} \, ,
  &
  \cos \Delta &= \frac{\mathcal{P}_{\mathcal{RI}}}{\sqrt{\mathcal{P}_{\mathcal{RR}}
  \mathcal{P}_{\mathcal{II}}}} \, .
\end{align}

Using Eqs.~\eqref{eq:zetaRY} and~\eqref{eq:zetaXY}, we can derive $\beta_{\rm iso}$ 
predicted by the entropy dilution mechanism, which depends on the initial perturbations, 
the dilution factor $\mathcal{S}$, and the composition of the final dark matter relic 
abundance $\Omega_{X_1}, \Omega_{X_2}$.
We have a rather simple result in the large $\mathcal{S}$ limit
\begin{align}
  \zeta_{R,f} &\simeq \zeta_{Y,i} \, , 
  &
  S_{XR} &\simeq 3 f_1\, (\zeta_{X,i}-\zeta_{Y,i}) \, ,
\end{align}
with $f_1$ defined in Eq.~\eqref{eq:f1f2}, and
\begin{equation}\label{eq:BetaIsoPrediction}
\begin{split}
  \beta_{\rm iso} &\simeq \frac{9 f_1^2(P_{XX}-2P_{XY}+P_{YY})}{P_{YY}+9 f_1^2(P_{XX}-2P_{XY}+P_{YY})} \, , 
  \\
  \cos\Delta &\simeq \frac{P_{XY}-P_{YY}}{\sqrt{P_{YY}(P_{XX}-2P_{XY}+P_{YY})}} \, ,
\end{split}
\end{equation}
where  the primordial power spectrum $P_{\alpha\beta} \sim \langle \zeta_{\alpha,i} 
\zeta_{\beta,i} \rangle$ is defined in terms of the initial perturbations, 
with $\alpha,\beta = X, Y, R$.
In the large $\mathcal{S}$ limit, the dependence on the power spectra related to $\zeta_{R, i}$ 
drops out of the result.
The positivity of $\mathcal{P}_{\mathcal{I}\mathcal{I}}$ implies the correlated spectrum 
$|P_{XY}| \leq \sqrt{P_{XX}P_{YY}}$.
The inequality is saturated in the $\zeta_{X, i} = \zeta_{Y,i}$ limit, where all perturbations 
in the universe are adiabatic to begin with, leading to $\beta_{\rm iso}=0$ and $\cos\Delta$ 
is ill-defined.
No isocurvature can be generated with adiabatic initial conditions~\cite{Weinberg:2004kr}.

In the presence of non-vanishing isocurvature initial perturbations, another possibility 
to suppress $\beta_{\rm iso}$ is to have $Y_X/Y_Y\ll n {\rm Br}_{Y\to X}$, {\it i.e.}, $f_1\ll 1$. 
By resorting to Eq.~\eqref{eq:OmegaXwithY}, this would imply that most of the dark matter 
$X$ in the universe today would have to arise from the dilutor $Y$ decay.

The Planck experiment gives the most stringent constraint on dark matter isocurvature~\cite{Planck:2018jri}. 
As an example to demonstrate the implications for the dark matter dilution mechanism 
discussed in this work, we consider the ``arbitrarily correlated'' case of primordial 
density perturbations, where $\zeta_X$ and $\zeta_Y$ originate from quantum 
fluctuations of two light scalars during inflation (one of them could be the inflaton).
We assume they have similar scale-invariant and uncorrelated power spectra $
P_{YY}=4P_{XX}$, $P_{XY}=0$. 
Using Eq.~\eqref{eq:BetaIsoPrediction} then gives
\begin{align}
  \beta_{\rm iso} \simeq \frac{45 f_1^2}{4+45 f_1^2} \, , & &
  \cos \Delta \simeq -0.89 \, .
\end{align}
The value of $\cos\Delta$ is very close to the fully anti-correlated case $\cos \Delta = -1$. 
We further assume that the two power spectra $\mathcal{P}_{\mathcal{R}\mathcal{R}}$ 
and $\mathcal{P}_{\mathcal{I}\mathcal{I}}$ have the same tilt.
This allows us to use the upper bound on $\beta_{\rm iso}$ corresponding to the 
row ``{\it curvaton II Planck TT,TE,EE+lowE+lensing}'' in Table 14 of~\cite{Planck:2018jri},
which in turn implies a strict upper bound of $f_1$
\begin{align}
  \beta_{\rm iso} &\lesssim 10^{-3} \, ,
  & &\Rightarrow &
  f_1 &< 0.0098 \, .
\end{align}
With Eq.~\eqref{eq:f1f2}, this result implies $f_2 \simeq 1$, {\it i.e.}, that
essentially the entire contribution of dark matter in the universe today must
come from the $X_2$ component sourced by the dilutor $Y$ decay.
The corresponding constraint is shown in Fig.~\ref{fig:moneyplot} by the orange 
excluded region in the ${\rm Br}_{Y\to X}$ versus $Y_X/Y_Y$ plane.

Note that $\cos \Delta$ is always negative if the initial perturbations in $X$ and 
$Y$ are uncorrelated and $P_{XY}=0$, see Eq.~\eqref{eq:BetaIsoPrediction}.
A future detection of positively-correlated dark matter isocurvature would
exclude this choice of initial conditions.

%
%
\section{Other Cosmological Probes} \label{sec:other}

In this section, we consider other cosmological constraints on the entropy dilution 
mechanism. 
Our discussion continues to assume that $X$ comprises the entirety of dark matter 
in the universe today; see Eq.~\eqref{eq:f1f2}.

\subsection{Free Streaming} \label{sec:freestreaming}

A characteristic outcome of the entropy dilution mechanism explored in this work is 
a universe with two component dark matter, $X_1, X_2$.
The fraction of $X_2$ dark matter is proportional to the dilutor $Y$ decay 
branching ratio ${\rm Br}_{Y\to X}$.
Due to the large $m_X$ versus $m_Y$ hierarchy, suggested by Eq.~\eqref{eq:OmegaXwithY},
the $X_2$ particles are ultra-relativistic after $Y$ decay, and this requires an
extended period of expansion of the universe for them to become non-relativistic.
We define $T_{\rm NR}$ as a temperature of background photons when their average
momentum falls below $m_X$.
One can derive
\begin{equation}\label{eq:TNR}
  T_{\rm NR} = T_* \left(\frac{m_X}{m_Y}\right) \left( \frac{g_*(T_*)}{g_*(T_{\rm NR})}
  \right)^{\frac{1}{3}}
  \simeq 0.7\,{\rm eV} \left( \frac{\Omega_Xh^2}{0.12} \right)\left( \frac{Y_X}{Y_Y} +
  n {\rm Br}_{Y\to X} \right)^{-1} \left( \frac{g_*(T_*)}{10.75} \right)^{\frac{1}{3}} \, ,
\end{equation}
where in the second step we applied Eq.~\eqref{eq:OmegaXwithY} and considered $T_{\rm NR}$ 
taking place after BBN with $g_*(T_{\rm NR}) = 3.9$. 
If the corresponding time is too late, free-streaming dampens the dark matter power spectrum 
and is constrained by large and small scale structure observations.

This has been considered in an early work by the authors by assuming 
$Y_X/Y_Y=1 \gg n {\rm Br}_{Y\to X}$~\cite{Nemevsek:2022anh}.
In this case, $T_{\rm NR}$ is fixed to the sub-eV scale, which is around the era of 
matter-radiation equality.
Fitting the large scale matter power spectrum measured by the SDSS 
experiment~\cite{Reid:2009xm} places an upper bound on 
$f_2 \simeq n {\rm Br}_{Y\to X} \lesssim 1 \%$.

\begin{figure}[t]
  \begin{center}
    \includegraphics[width=0.618\textwidth]{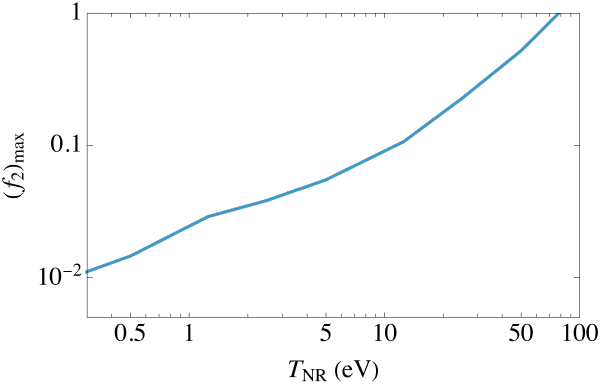}
  \end{center}
  \caption{Upper bound on the secondary $X$ contribution to the dark matter relic as a function
  of $T_{\rm NR}$, which follows from a similar analysis of the large-scale structure constraint 
  in~\cite{Nemevsek:2022anh, Nemevsek:2023yjl}. }
  \label{fig:X2fraction}
\end{figure}

In this work, we go beyond the previous assumption and explore smaller values of 
the $Y_X/Y_Y$ ratio. 
Eq.~\eqref{eq:TNR} tells us that by reducing the $Y_X/Y_Y$ ratio, the temperature 
$T_{\rm NR}$ can be increased to weaken the structure formation constraints on 
$n{\rm Br}_{Y\to X}$.
We repeat the analysis in~\cite{Nemevsek:2022anh, Nemevsek:2023yjl} by using the same 
primordial phase space distribution functions for $X_2$, but allow $T_{\rm NR}$ to vary 
as a free parameter.
The resulting upper bound on $f_2$ is depicted in Fig.~\ref{fig:X2fraction}.
A similar upper bound is obtained by applying the CMB constraint on $C_\ell^{TT}$~\cite{
Xu:2021rwg}.~\footnote{Yu-Ming Chen, private communication.}
We apply this upper bound on the $Y_X/Y_Y$ versus $n{\rm Br}_{Y\to X}$ parameter space, 
which excludes the blue shaded region in Fig.~\ref{fig:moneyplot}.
For clarity, we consider the case with $n=2$ and $y=1$, which corresponds to the decay 
mode $Y\to XX$.
The structure formation constraint goes away for $T_{\rm NR}>80\,$eV, because the dark
matter free-streaming length is below the smallest scales probed by the SDSS and 
Lyman-$\alpha$ data.
This is the reason why the blue curve plateaus for very small $Y_X/Y_Y$.

The structure formation constraint and the one from dark matter isocurvature 
(discussed in the previous section) are complementary to each other.
In the presence of sizable primordial isocurvature perturbations, they together require
$Y_X/Y_Y\lesssim 10^{-4}$, which in turn implies that $Y\to X$ decay must contribute 
the majority of the dark matter relic density.

It is worth emphasizing that these constraints are surprisingly robust, as noticed 
in~\cite{Nemevsek:2022anh}.
They are essentially independent of the detailed model parameters, such as $m_X$, $m_Y$,
and $\tau_Y$, as long as $X$ comprises the entirety of dark matter in the universe. 

\subsection{$\Delta N_{\rm eff}$}\label{sec:neff}

Due to the relatively low $T_{\rm NR}$ found in Eq.~\eqref{eq:TNR}, the secondary $X$
particles from $Y$ decay are still relativistic during BBN and act as an additional
radiation degree of freedom, usually parametrized by $\Delta N_{\rm eff}$. 
The contribution to $\Delta N_{\rm eff}$ here is controlled by the $y {\rm Br}_{Y\to X}$ 
combination. 
Immediately after $Y$ decay, the energy density ratio between the secondary $X$ 
particles and the SM sector is given by 
\begin{equation}
  \left.\frac{\rho_{X_2}}{\rho_{\rm SM}}\right|_{T_*} = \frac{y {\rm Br}_{Y\to X}}{
  1-y {\rm Br}_{Y\to X}} \, .
\end{equation}
Afterwards, both evolve as radiation before the universe cools to $T_{\rm NR}$. 
The $X$ particles remain non-interacting, but the SM sector slightly ``heats up'' 
due to interactions and the change of $g_*$.
Around MeV temperatures, the ratio becomes
\begin{equation}
  \left.\frac{\rho_{X_2}}{\rho_{\rm SM}}\right|_{\rm MeV} = \frac{y {\rm Br}_{Y\to X}
  }{1-y {\rm Br}_{Y\to X}} \left( \frac{43}{4g_*(T_*)} \right)^{1/3} \, .
\end{equation}
This corresponds to~\cite{Nemevsek:2022anh}
\begin{equation}
  \Delta N_{\rm eff} \simeq \frac{43}{7} \frac{y {\rm Br}_{Y\to X}}{1-y {\rm Br}_{Y\to X}}
  \left( \frac{43}{4g_*(T_*)} \right)^{1/3} \ ,
\end{equation}
which holds at temperatures $T > T_{\rm NR}$.
In Fig.~\ref{fig:moneyplot}, the purple shaded region is excluded by requiring 
$\Delta N_{\rm eff} < 0.42$~\cite{Pitrou:2018cgg}. 
We set $y = 1$ and $g_*(T_*) = 10.75$, corresponding to the strongest possible constraint one 
could obtain. 
Even in this case, it is subdominant to the large-scale structure constraint (blue) unless 
$Y_X/Y_Y\gg1$.

%
%
\section{Conclusions and Outlook}

In the entropy dilution mechanism for addressing the origin of dark matter, the early universe 
undergoes a stage of temporary matter domination by a non-relativistic fluid made of particles $Y$.
The universe returns to regular radiation domination prior to BBN because $Y$ mainly decays into SM 
particles and injects entropy into the thermal plasma.
The decay of $Y$ can also produce a fraction of dark matter $X$, on top of a population of $X$ that 
predates the $Y$ domination era.
We explore possible signatures of such cosmological scenarios in a three-fluid ($X$, $Y$, SM radiation) 
setup by allowing for generic initial conditions in their primordial energy densities and perturbations.
We derive state-of-the-art constraints on dark matter isocurvature, free-streaming, and contribution to 
$\Delta N_{\rm eff}$ using CMB and large scale structure data.
These constraints are set on two key parameters of the entropy dilution mechanism: the dilutor-to-dark-matter 
decay branching ratio ${\rm Br}_{Y\to X}$ and the ratio of primordial populations $Y_X/Y_Y$.
Remarkably, our result is independent of the detailed model parameters, such as the mass of dark 
matter $m_X$ and the mass and lifetime of the dilutor $m_Y, \tau_Y$. 
Future experiments like LiteBIRD~\cite{LiteBIRD:2022cnt} and DESI~\cite{Ravoux:2023bgw, 
Karacayli:2023afs} can explore predictions of the dilution mechanism with higher precision.

Our work makes several contributions to this topic.
We present a conceptually new derivation of the entropy dilution factor by taking advantage of a 
non-standard scaling window for radiation right before $Y$ decay, where the radiation energy density 
evolves as the scale factor to the $(-3/2)$ power. 
Based on this finding, we evolve the large scale perturbations relevant for observations, which are 
super-horizon modes during $Y$ domination. 
In the same scaling window, the density perturbation of the SM plasma is driven towards that of $Y$,
and their difference goes as the entropy dilution factor to the $(-4/3)$ power, in stark
contrast to the usual freeze-out exponential dependence.
Finally, the density perturbation in the dark matter sector requires more care.
The secondary $X$ particles from $Y$ decay are energetic and could take a longer period to become 
non-relativistic compared to the primordial $X$ population.
This renders the possibility that dark matter isocurvature vanishes immediately after $Y$ decay
but re-emerges at later times.
To have a viable cosmology, either the initial perturbations of $X, Y$ are adiabatic, or the primordial 
$X$ must comprise a tiny fraction of the dark matter relic density today.

Some aspects of the dilution mechanism require further investigation. 
If the early matter domination lasts for a sufficiently large number of $e$-folds, non-relativistic 
dark matter perturbations with wavelengths below the horizon size could grow (linearly) into large values,
giving rise to rich phenomenology such as dark matter mini-halo formation~\cite{Erickcek:2011us, Blinov:2021axd},
gravitational wave signals~\cite{Baumann:2007zm, Fernandez:2023ddy}, or even primordial black holes~\cite{Clark:2016nst}. 
In the scenarios we consider, this could only happen for the primordial $X$ particles if they 
are born cold and isocurvature free, but not for the secondary $X$ particles from the dilutor decay.
Our discussion also did not address the origin of the cosmic baryon asymmetry. 
In the dilution mechanism, any pre-existing baryon asymmetry is diluted by a factor of $\mathcal{S}$.
In order to avoid overly large isocurvature in the baryon sector, the asymmetry had 
better be generated after the entropy dilution~\cite{Lyth:2002my, Moroi:2002rd}. 
This could place a further constraint on the temperature $T_*$ immediately after $Y$ decay 
in order to accommodate baryogenesis mechanisms that rely on electroweak sphalerons.

\section*{Acknowledgment}

We thank Yu-Ming Chen for useful discussions.
MN is supported by the Slovenian Research Agency under research core funding 
No. P1-0035 and in part by the research grant N1-0253 and the ARIS Fokus grant.
MN would also like to thank the CERN theory department for its hospitality during 
the completion of this work.
YZ is supported by a Subatomic Physics Discovery Grant (individual) from the 
Natural Sciences and Engineering Research Council of Canada.

\appendix
\section{Evolution Equation for $\zeta$}\label{app:zetatot}

In this section, we present some details of the derivation of the $\zeta$ evolution equation 
Eq.~\eqref{eq:ZetaTotalEvolve}, where $\zeta$ is defined in Eq.~\eqref{eq:ZetaTotal} as
\begin{equation}\label{appeq:zeta}
  \zeta = \frac{\sum_\alpha \dot{\bar\rho}_\alpha \zeta_\alpha}{\dot{\bar\rho}} 
  = \frac{\sum_\alpha {\bar\rho}_\alpha' \zeta_\alpha}{{\bar\rho}'} \, ,
\end{equation}
where $\bar\rho = \bar\rho_Y + \bar\rho_R + \bar\rho_X$, and the sum goes over $\alpha = Y, R, X$.
Because $\zeta$ is defined in terms of $\bar\rho_\alpha$ and $\zeta_\alpha$, its evolution is jointly 
governed by Eqs.~\eqref{eq:RhoYEvol}--\eqref{eq:RhoXEvol} and \eqref{eq:zetaY}--\eqref{eq:zetaX}. 

First, the average energy densities satisfy
\begin{equation}\label{appeq:A2}
\begin{split}
  \bar\rho_\alpha' &= -3 (1+w_\alpha) \bar\rho_\alpha + \frac{Q_\alpha}{H} \, , 
  \\
  \bar\rho_\alpha'' &= -3 (1+w_\alpha) \bar\rho_\alpha' + \frac{Q_\alpha}{H} 
  \left( \frac{\bar\rho_Y'}{\bar\rho_Y} - \frac{H'}{H} \right) \, ,
\end{split}
\end{equation}
where $Q_\alpha$ is given in Eq.~\eqref{eq:QandDeltaQ} and we assume $w_\alpha$ to be constant. 
An important property is the net zero of heat transfer
\begin{equation}
  \sum_\alpha Q_\alpha = 0 \, ,
\end{equation}
which allows us to obtain
\begin{equation}\label{appeq:rho''}
  \bar\rho'' = -3 \sum_\alpha \left(1 + w_\alpha \right) \bar \rho_\alpha' = -4\bar\rho' + \bar\rho_Y' \, .
\end{equation}
In the last step, we used the equation of state $w_Y = 0$ and for all its decay products $w_R = w_X = 1/3$.

Next, we differentiate Eq.~\eqref{appeq:zeta}
\begin{equation}\label{appeq:A5}
\begin{split}
\zeta' &= \frac{\left(\sum_\alpha {\bar\rho}_\alpha' \zeta_\alpha \right)'}{{\bar\rho}'} - \frac{\left(\sum_\alpha {\bar\rho}_\alpha' \zeta_\alpha \right) \bar\rho''}{{\bar\rho}'^2}  \\
&= \frac{\left(\sum_\alpha {\bar\rho}_\alpha' \zeta_\alpha \right)'}{{\bar\rho}'} + \frac{4 \left(\sum_\alpha {\bar\rho}_\alpha' \zeta_\alpha \right)}{{\bar\rho}'} - \frac{\left(\sum_\alpha {\bar\rho}_\alpha' \zeta_\alpha \right) \bar\rho_Y'}{{\bar\rho}'^2} \\
&= \frac{\left(\sum_\alpha {\bar\rho}_\alpha' \zeta_\alpha \right)'}{{\bar\rho}'} + \frac{4 \left(\sum_\alpha {\bar\rho}_\alpha' \zeta_\alpha \right)}{{\bar\rho}'} - \frac{\zeta \bar\rho_Y'}{{\bar\rho}'} \\
&= \frac{\left(\sum_\alpha {\bar\rho}_\alpha' \zeta_\alpha' \right)}{{\bar\rho}'} + \frac{\left(\sum_\alpha {\bar\rho}_\alpha'' \zeta_\alpha \right)}{{\bar\rho}'} + \frac{4 \left(\sum_\alpha {\bar\rho}_\alpha' \zeta_\alpha \right)}{{\bar\rho}'} - \frac{\zeta \bar\rho_Y'}{{\bar\rho}'} \ ,
\end{split}
\end{equation}
where we apply Eq.~\eqref{appeq:rho''} in the second step and Eq.~\eqref{appeq:zeta} in the third step.

Using Eq.~\eqref{appeq:A2} again, we derive
\begin{equation}
  \sum_\alpha {\bar\rho}_\alpha'' \zeta_\alpha = - 4 \left( \sum_\alpha \bar\rho_\alpha' 
  \zeta_\alpha \right) + \bar\rho_Y' \zeta_Y + \frac{1}{H} \left( \frac{\bar\rho_Y'}{\bar\rho_Y} 
  - \frac{H'}{H} \right) \sum_\alpha Q_\alpha \zeta_\alpha \, .
\end{equation}
It simplifies Eq.~\eqref{appeq:A5} to
\begin{equation}\label{appeq:A7}
  \zeta' = \frac{\left(\sum_\alpha {\bar\rho}_\alpha' \zeta_\alpha' \right)}{
  {\bar\rho}'} + \frac{\bar\rho_Y' \zeta_Y}{\bar\rho'} + \frac{1}{H \bar\rho'} 
  \left( \frac{\bar\rho_Y'}{\bar\rho_Y} - \frac{H'}{H} \right) \sum_\alpha Q_\alpha 
  \zeta_\alpha - \frac{\zeta \bar\rho_Y'}{{\bar\rho}'} \, .
\end{equation}

For the first term on the right-hand side, we use Eqs.~\eqref{eq:zetaY}--\eqref{eq:zetaX}, 
which can be written in the following general form for any $\alpha$
\begin{equation}
\bar\rho_\alpha'\zeta_\alpha' = \frac{Q_\alpha H'}{H^2} (\zeta_\alpha - \zeta) - 
\frac{Q_\alpha \bar\rho_Y'}{\bar\rho_Y H} (\zeta_\alpha - \zeta_Y) \, .
\end{equation}
Summing over $\alpha$ and using $\sum_\alpha Q_\alpha=0$ gives
\begin{equation}
\sum_\alpha \bar\rho_\alpha'\zeta_\alpha' = \frac{1}{H} \left(\frac{H'}{H} - \frac{\bar\rho_Y'}{\bar\rho_Y} \right) \sum_\alpha Q_\alpha \zeta_\alpha \ .
\end{equation}
As a result, on the right-hand side of Eq.~\eqref{appeq:A7}, the first term cancels the third term, yielding
\begin{equation}
\zeta' = - \frac{\bar\rho_Y'}{\bar\rho'} (\zeta - \zeta_Y) \ ,
\end{equation}
which is the same as Eq.~\eqref{eq:ZetaTotalEvolve}.

\section{Conservation of $\zeta_{X}$ When Dark Matter Turns Non-relativistic}\label{app:zetaconservation}

In our discussion, the dark matter particles are relativistic at an early time and later become
cold due to the expansion of the universe.
The radiation-to-matter transition is described by a continuous change of $w_X$ from $1/3$ to $0$.
In this appendix, we shall show that the corresponding perturbation $\zeta_{X}$ is invariant during the transition.
The same conclusion holds for $X_1, X_2$ particles; thus we leave out the subscript in the discussion.

In the absence of a source term, the continuity equations of $X$ are (see Eq.~\eqref{eq:RhoDeltaRhoEvolution})
\begin{equation}\label{appeq:B1}
\begin{split}
    \dot{\bar\rho}_X + 3 H (1+w_X) \bar\rho_X &= 0 \ , \\
    \delta\dot{\rho}_X + 3 H\left( 1 + \frac{\delta P_X}{\delta \rho_X} \right) \delta \rho_X - 3 \dot\phi (1+w_X) \bar\rho_X &= 0 \ ,
\end{split}
\end{equation}
where $w_X=\bar P_X/\bar\rho_X$.

We assume the $w_X=1/3\to0$ transition takes place well after $Y$ decay and when the universe is still radiation dominated.
The first thing we want to show is that the metric perturbation $\phi$ is conserved on super-horizon scales.
Using the Einstein equation and the continuity equation for radiation ($w_R=1/3$)
\begin{equation}
\begin{split}
3 H\dot\phi + 3H^2 \phi = - 4\pi G \delta \rho_R &= - \frac{3H^2}{2} \frac{\delta \rho_R}{\bar\rho_R} \ , \\
    \dot{\bar\rho}_R + 4 H \bar\rho_R &= 0 \ , \\
    \delta\dot{\rho}_R + 4 H \delta \rho_R - 4 \dot\phi \bar\rho_R &= 0 \ ,
\end{split}
\end{equation}
as well as $H=1/(2t)$ valid for a radiation dominated universe, we can obtain a second-order differential equation for $\phi$,
\begin{equation}
2 t \ddot\phi + 5 \dot \phi = 0 \ . \\
\end{equation}
The solution is given by $\phi \sim A + B t^{-3/2}$. Neglecting the damping mode, $\phi$ is a constant of time $t$.

Setting $\dot\phi=0$, Eq.~\eqref{appeq:B1} reads
\begin{equation}\label{appeq:B4}
\begin{split}
    \dot{\bar\rho}_X + 3 H (1+w_X) \bar\rho_X &= 0 \, , 
    \\
    \delta\dot{\rho}_X + 3 H\left( 1 + \frac{\delta P_X}{\delta \rho_X} \right) \delta \rho_X &= 0 \, .
\end{split}
\end{equation}

For the $\delta P_X/\delta\rho_X$ factor, we apply the adiabatic relation
\begin{equation}
\frac{\delta P_X}{\delta \rho_X} = \frac{\dot{\bar{P}}_X}{\dot{\bar{\rho}}_X} = \frac{w_X \dot{\bar{\rho}}_X + \dot w_X \bar\rho_X}{\dot{\bar{\rho}}_X} = w_X - \frac{\dot w_X}{3H(1+w_X)} \ .
\end{equation}
In the last step, we use the first equation of \eqref{appeq:B4}.
With this, the second equation of \eqref{appeq:B4} can be written as
\begin{equation}
\frac{d}{dt} \ln \left( \frac{\delta \rho_X}{\bar\rho_X} \right) = \frac{d}{dt} \ln (1+w_X) \ .
\end{equation}
Integrating from the time when $X$ was relativistic to the time when $X$ became cold, we find 
\begin{equation}
\left( \frac{\delta \rho_X}{(1+w_X)\bar\rho_X} \right)_< = \left( \frac{\delta \rho_X}{(1+w_X)\bar\rho_X} \right)_> \ ,
\end{equation}
or
\begin{equation}
\left( H \frac{\delta \rho_X}{\dot{\bar\rho}_X} \right)_< = \left( H \frac{\delta \rho_X}{\dot{\bar\rho}_X} \right)_> \ .
\end{equation}

This result, together with $\dot\psi \simeq \dot\phi=0$ found earlier, implies that
\begin{equation}
\zeta_X \equiv -\psi - H \frac{\delta \rho_X}{\dot{\bar\rho}_X} \ ,
\end{equation}
is a conserved quantity during the radiation-to-matter transition of $X$ particles.

For slowly varying $w_X$, the conservation of $\zeta_X$ holds only for super-horizon modes. 
In~\cite{Nemevsek:2025zyo}, it has been further proven that $\zeta_X$ is preserved for modes 
of all wavelengths if a first-order phase transition takes place and rapidly changes $w_X$ 
from $1/3$ to zero.

\bibliographystyle{apsrev4-1}
\bibliography{refs}{}
\end{document}